\documentclass[prx,twocolumn,longbibliography,superscriptaddress,preprintnumbers]{revtex4-2}
\usepackage[colorlinks,bookmarks=true,citecolor=blue,linkcolor=blue,urlcolor=blue]{hyperref}
\usepackage{dcolumn,graphicx,amsfonts,amsthm,bm,color,appendix,float,tabularx}
\usepackage{braket}
\usepackage[normalem]{ulem}
\usepackage[version=4]{mhchem}
\usepackage{siunitx}
\usepackage{amsmath}
\usepackage{amssymb}
\usepackage{CJKutf8}
\usepackage{multirow}
\usepackage{booktabs}
\usepackage{comment}
\usepackage[commandnameprefix=ifneeded]{changes}
\usepackage{hyperref}

\usepackage[dvipsnames]{xcolor}
\definecolor{pal0}{rgb}{0.8941, 0.102 , 0.1098}
\definecolor{pal1}{rgb}{0.2157, 0.4941, 0.7216}
\definecolor{pal2}{rgb}{0.302 , 0.6863, 0.2902}
\definecolor{pal3}{rgb}{0.5961, 0.3059, 0.6392}
\definecolor{pal4}{rgb}{1.    , 0.498 , 0.    }

\renewcommand{\v}[1]{\boldsymbol{#1}}
\renewcommand{\Im}{\operatorname{Im}}

\newcommand{\PRLsec}[1]{\emph{#1---}}

\usepackage{enumitem,amssymb}
\newlist{todolist}{itemize}{2}
\setlist[todolist]{label=$\square$}
\usepackage{pifont}
\usepackage{blindtext}

\begin{document}

\title{Large scale neural quantum states reveal the interplay between superconductivity and quantum criticality in the Hofstadter-Hubbard model}
\author{Christopher Roth}
\affiliation{Center for Computational Quantum Physics, Flatiron Institute, 162 5th Avenue, New York, NY 10010, USA}

\author{Andrew Millis}
\affiliation{Center for Computational Quantum Physics, Flatiron Institute,
162 5th Avenue, New York, NY 10010, USA}
\affiliation{Department of Physics, Columbia University, New York, NY 10027, USA}

\author{Tomohiro Soejima (\begin{CJK*}{UTF8}{bsmi}副島智大\end{CJK*})}
\affiliation{Center for Computational Quantum Physics, Flatiron Institute,
162 5th Avenue, New York, NY 10010, USA}
\affiliation{Center for Quantum Phenomena, Department of Physics,
New York University, 726 Broadway, New York, New York 10003, USA}

\begin{abstract}
Understanding how a parent insulating state shapes the superconductivity that emerges upon doping is a long-standing problem dating back to Anderson's resonating-valence-bond proposal. The triangular-lattice Hofstadter-Hubbard model with $\pi/2$ flux per plaquette offers an ideal setting: at half filling it hosts two distinct parent states---an integer quantum Hall insulator and a chiral spin liquid---separated by a topological phase transition. Using neural quantum states on tori of up to $432$ sites, we present strong evidence that the transition is continuous, with a vanishing $2e$ charge gap and critical charge fluctuations.
Upon doping, we find a topological superconductor on either side of the transition.
While the pairing order parameter remains nearly unchanged across the transition, the superfluid stiffness is strongly enhanced near the critical point.
The energy scale of the superconductor is therefore set not by which parent state is doped, but by proximity to the transition between them.
Our results establish neural quantum states as a powerful tool for understanding the interplay between unconventional electronic correlations and superconductivity.

\end{abstract}

\maketitle

In the early days following the discovery of high transition-temperature superconductivity in the copper-oxide materials, P.W. Anderson argued that there existed a class of remarkable insulating states, in which electrons were paired in such a way that if carrier motion were released by adding or removing electrons, superconductivity would result \cite{AndersonRVB}. Subsequent work has added substantial theoretical perspective to Anderson's original idea, including that the insulating states he intuited could be understood as quantum spin liquids  \cite{laughlin_relationship_1988, savary2017quantum, zhou2017quantum, wen2017colloquium, lee2006doping, song2021doping}.
Despite this theoretical progress, there is limited knowledge about how the parent state affects the energy scales associated with superconductivity and thus influences the transition temperature. 

An ideal model for understanding this interplay is realized by the Hofstadter-Hubbard model on the triangular lattice with $\pi/2$ flux around each triangular plaquette. 
It is believed to host several different parent states at half filling---an integer quantum Hall state (IQH) with a band gap when the on-site interaction is weak, and a chiral spin liquid (CSL)~\cite{kalmeyer1987equivalence} at intermediate coupling \cite{kuhlenkamp2024chiral,Divic2026PRB, gallegos2026PRB}, and at very large $U$ a $120^\circ$ antiferromagnet \cite{macdonald1988prb,viteritti2026approaching}.
Numerical studies suggest the phase transition between IQH and CSL is second order. The difference in Hall conductivity between the IQH and CSL phases implies the existence of gapless charged excitations at  this critical point and a  field theoretical analysis suggests the emergence of a topological superconductor upon doping~\cite{Divic2025PNAS}.

Existing numerical studies on this problem find pairing of charged excitations~\cite{Divic2025PNAS}, quasi-long-range order of the superconducting order parameter in quasi-1D~\cite{Kuhlenkamp2025Arxiv, Chen2026PRL}, large pairing susceptibility at low temperature~\cite{Chen2026PRL}, and nonzero superconducting order parameter~\cite{niu2025thermodynamic}. However, the small system sizes or correlation lengths these methods can reach has limited their ability to access quantum critical physics,  such as the existence of gapless excitations. Furthermore, these calculations have not estimated how the superconducting energy scales change throughout the phase diagram. 

In this work, we use neural quantum states~\cite{carleo2017solving, spencer2020better,nomura2021dirac, lange2024architectures} to accurately simulate large systems of up to $432$ sites.
First, we resolve the quantum critical point between the IQH and CSL phases. We show that very large system sizes, accessible to date only with NQS methods, are required to observe the theoretically predicted vanishing of the $2e$ charge gap and demonstrate the transition point is associated with critical charge fluctuations. 
Upon doping, we find a topological superconducting phase with off-diagonal long-range order (ODLRO) and nontrivial spin quantum Hall conductance.
Near the critical point, charge fluctuations strongly enhance the superfluid stiffness, leading to increased superconducting coherence. In contrast, the pairing order parameter is largely unaffected by proximity to the critical point, suggesting the pair formation is not enhanced by quantum criticality.
Thus, the behavior of the superconductor is governed less by which parent insulator is doped than by proximity to the transition between them.
Our work establishes NQS as a uniquely viable tool for capturing phases with two-dimensional long-correlation-length fluctuations and resolving the interplay between superconductivity and quantum criticality.

\begin{figure}[t] 
    \centering
    \includegraphics[width=1.\linewidth]{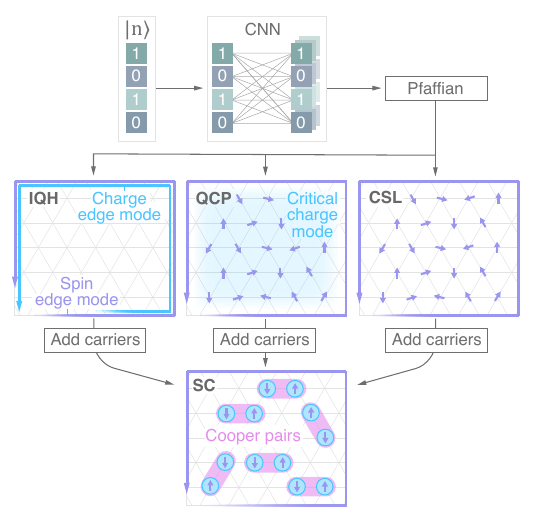}
    \caption{Overview of our numerical simulations. We use a Pfaffian-based neural quantum wavefunction to simulate the triangular-lattice Hofstadter-Hubbard model at $\pi/2$ flux per plaquette. We find a second-order phase transition between an IQH and a CSL phase at half filling, separated by a quantum critical point. Upon doping, all of these phases become the same superconductor, but the superfluid stiffness is enhanced near the critical point.} 
    \label{fig: overview}
\end{figure}

\PRLsec{Hamiltonian}
We consider the triangular-lattice Hofstadter-Hubbard model with $\Phi = \pi/2$ flux through each plaquette. We can choose a gauge such that all hoppings are imaginary:
\begin{equation}
    H  = it\sum_{i,j} \tau_{i, j} c^\dagger_i c_j + \sum_i U_i \Big(n_{i\uparrow} - \frac{1}{2}\Big)\Big( n_{i\downarrow} - \frac{1}{2}\Big),
    \label{eq:flux_hofstadter_hubbard}
\end{equation}
where $\tau_{ij} = \pm 1$. The purely imaginary hoppings endow the model with $SO(4)$ symmetry at half filling~\cite{Divic2026PRB, Divic2025PNAS}, generated from spin $SU(2)$ symmetry and charge $SU(2)$ symmetry~\cite{Yang1989,1990YangZhang,Yang1991,Affleck1988zj}. The Hamiltonian is particle-hole symmetric about half filling, so electron and hole dopings are equivalent. The details of the model are shown in App.~\ref{sec: appendix geometry+gauge}. 
At $U=0$, the ground state at half filling is an integer quantum Hall state with $C=1$ per spin ($C_{\rm tot}=2$). As mentioned above, upon increasing $U$, the model exhibits an IQH--CSL phase transition~\cite{kuhlenkamp2024chiral,Divic2026PRB, gallegos2026PRB}, and doping it can give rise to superconductivity~\cite{Divic2025PNAS, Kuhlenkamp2025Arxiv, Chen2026PRL, niu2025thermodynamic}. 

\PRLsec{Hidden Fermion Pfaffian States}
We train a variational wavefunction with the Hidden Fermion Pfaffian State (HFPS) architecture \cite{chen2025neural} to find the ground state of the Hofstadter-Hubbard model. This wavefunction, which introduces hidden fermions in order to learn correlation on top of a mean field fermion state, has been used to understand superconductivity in the $t-t^\prime$ square lattice Hubbard model \cite{roth2025superconductivity} as well as t-J bilayers \cite{lange2026simulating}. Here we use a residual convolutional neural network architecture \cite{chen2024empowering} to parameterize the couplings between electrons and hidden fermions. The pairing orbitals for the visible fermions are chosen to be block-circulant such that the wavefunction has the underlying translational symmetry of the Hamiltonian. Therefore, we can compute a translationally symmetric wavefunction with only a single Pfaffian evaluation, and we can efficiently scale to large system sizes. We consider $C_6$ symmetric clusters of size $\sqrt{3}L \times \sqrt{3} L$ with periodic boundary conditions. Simulations were performed using the \texttt{Quantax}~\cite{quantax} package, which uses \texttt{lrux}~\cite{lrux} to accelerate the computation of determinants and pfaffians. The network and training procedure are described in detail in App.~\ref{sec: appendix hyperparameters}.

\PRLsec{Vanishing charge gap}
\begin{figure}[t]
    \centering
    \includegraphics[width=\linewidth]{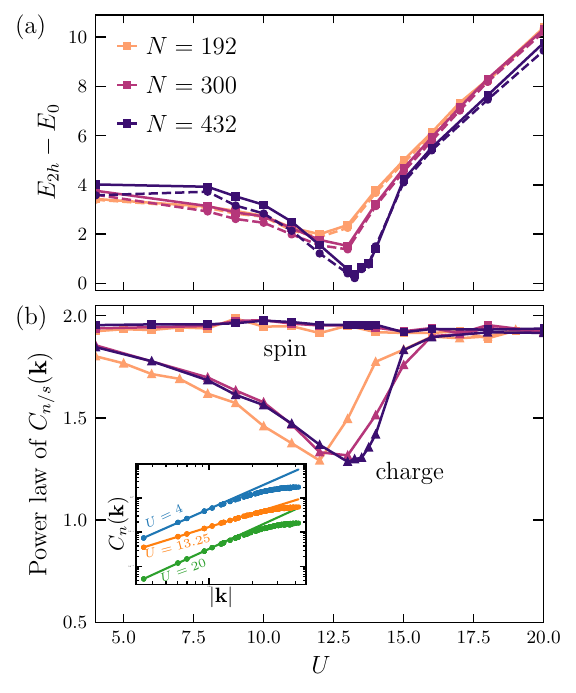}
    \caption{Second-order phase transition between the IQH and CSL phases. Here we show the results of simulations on $N=192$, $N=300$, and $N=432$ site clusters. (a) Energy gap to a 2 hole excitation as a function of $U$. The charge gap approaches zero around $U\sim 13.25$. (b) Small-$|{\bf k}|$ power law of the charge (triangle) and spin (square) structure factors as a function of $U$. Charge fluctuations become enhanced near the critical point. (Inset) Power-law fit of $C_{\hat n}(\v{k})$ at $U=4, 13.25,$ and $20$.
    }   
    \label{fig: main QCP}
\end{figure}
\begin{figure}[t]
    \centering
    \includegraphics[width=\linewidth]{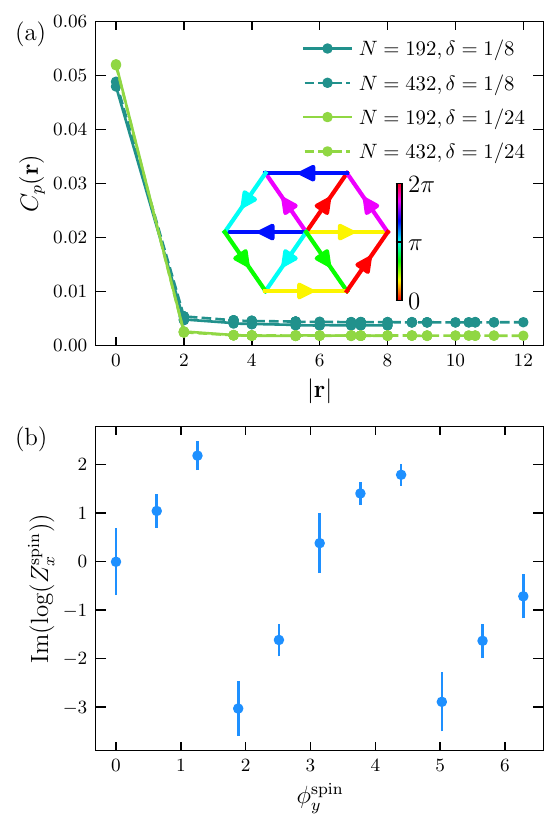}
    \caption{Characterizing topological superconductivity. (a) Pair correlations on the Hofstadter-Hubbard model at $U=8$ for $192$ and $432$ site clusters at $1/8$ and $1/24$ doping. (Inset) The phases of the pairing order parameter as written in Eq.~\eqref{eqn: order param definition}.  (b) Many-body spin polarization as a function of spin flux threading at $U=10, \delta = 1/36$ on the $192$ site torus. The polarization winds twice as $\phi_y^{\rm spin}$ goes from $0$ to $2\pi$, corresponding to $\sigma_{xy}^s = 2 \frac{\hbar}{8\pi}$. }
    \label{fig: topoSC}
\end{figure}
We first show that our variational ansatz can correctly capture the two gapped phases. We use the following flux-threading procedure to measure the Hall conductance and the spin quantum Hall conductance~\cite{zaletel2013topological,kuhlenkamp2024chiral, gallegos2026spinless}. We find the variational ground state at $U = 10$ and $U=16$ at different values of charge and spin flux. By computing the evolution of many-body polarization with charge/spin flux, we obtain $\sigma_{xy}^c = 2\frac{e^2}{h}, \sigma_{xy}^s = 2 \frac{\hbar}{8\pi}$ at $U= 10$, and $\sigma_{xy}^c = 0, \sigma_{xy}^s = 2 \frac{\hbar}{8\pi}$ at $U = 16$, consistent with IQH and CSL phases, respectively (see App.~\ref{sec: appendix IQH-CSL}).

By tuning $U$ to the critical point, our variational ansatz shows a vanishing $2e$ charge gap. In Fig.~\ref{fig: main QCP}(a) we show the $2e$ charge gap at different system sizes. The solid lines show our estimate of the $2e$ charge gap using our variational wavefunctions at $N_e = N_{\mathrm{site}}$ and $N_e = N_{\mathrm{site}} - 2$ and the dashed lines show the result of a zero-variance extrapolation as described in App.~\ref{sec: appendix zero-variance}. 
The difference between the raw and extrapolated energies is very small, showing that our variational ansatz likely captures the energy gap to good accuracy.  

As system size increases the minimum value of the $2e$ gap trends toward zero and shifts from $U \approx 12$ to $U \approx 13$. On the $432$ site cluster, the smallest gap is at $U=13.25$, and has a value of $0.22 t$, roughly a factor of $20$ smaller than the band gap.
This is in contrast to an earlier DMRG study that only found a $\sim 25 \%$ decrease relative to the band gap \cite{Divic2025PNAS}.

\PRLsec{Structure factor}
The small-$|\v{k}|$ behavior of the structure factor further reveals the nature of the critical point. Let us decompose the correlation function into an analytic part and a power-law part $C(x) = C_{\rm analytic}(x) + c|x|^{-\eta}$. Its Fourier transform at small $|k|$ takes the form $C(k) = \alpha |k|^2 + \beta |k|^{\eta - 2}$ with a log correction when $\eta = 4$. Nonanalyticity of $C(k)$ as $|k|\to 0$ therefore signals power-law decay of the correlation function. We consider the charge and spin structure factors $C_{\hat n}({\bf k})$ and $C_{\hat S^z}({\bf k})$ (see App.~\ref{sec: structure factors}), and fit the low-$|\v{k}|$ data to a power law (Fig.~\ref{fig: main QCP}(b)). While the spin structure factor always scales quadratically at small $|{\bf k}|$, the charge structure factor has a much smaller scaling exponent near the critical point, signaling critical correlations.

\PRLsec{Superconductivity upon doping} 
We next find that doping the Hofstadter-Hubbard model results in a topological superconductor, in agreement with theoretical and numerical predictions~\cite{Divic2025PNAS, kuhlenkamp2024chiral, Chen2026PRL}.

To capture superconductivity, we first establish the existence of off-diagonal long-range order. We consider nearest neighbor pairing of the form
\begin{equation} \label{eqn: order param definition}
 \Delta_{\v{x}} = \frac{1}{12 \sqrt{2}} \sum_{\langle \alpha, \beta\rangle \in U_{\v{x}} } \textrm{phase}(\alpha, \beta) (c_{\alpha, \uparrow} c_{\beta, \downarrow} - c_{\alpha, \downarrow} c_{\beta, \uparrow} )
\end{equation}
where $\langle \alpha, \beta\rangle \in U_{\v{x}}$ includes all nearest neighbor bonds in a single magnetic unit cell centered at ${\bf x}$, and the phases of these bonds are shown in the inset to Fig.~\ref{fig: topoSC}(a). Here, as in Refs.~\cite{Divic2025PNAS, Kuhlenkamp2025Arxiv, Chen2026PRL, niu2025thermodynamic}, we find the dominant pairing mode has $2 \pi$ winding about the rotational centers. ODLRO can be diagnosed by the correlator
\begin{equation}
C_p(\v{r}) =  \frac{1}{N} \sum_{\bf x} \langle \Delta^\dagger_{\bf r + \bf x} \Delta_{\bf x} \rangle.
\end{equation}
In Fig.~\ref{fig: topoSC}(a), we show the pair correlations $C_p(\v{r})$ at $U=8$ for different values of doping. The pair correlation function plateaus to a constant, indicating ODLRO.

The topological nature of the superconductor can be elucidated from spin quantum Hall conductivity. We compute this from flux-threading, analogously to the computation for IQH and CSL. In Fig.~\ref{fig: topoSC}(b), we show the evolution of the many-body spin polarization $Z_x^{\rm spin} = \exp[\frac{4\pi i}{3L} (\hat{X}_\uparrow - \hat{X}_\downarrow)]$ as we thread spin flux through the system. The polarization winds twice, consistent with spin quantum Hall conductivity $\sigma_{xy}^s = 2 \frac{\hbar}{8\pi}$, in agreement with topological superconductivity predicted in the literature~\cite{Divic2025PNAS, Kuhlenkamp2025Arxiv, Chen2026PRL, niu2025thermodynamic}.

\PRLsec{Interplay of criticality and superconductivity}
\label{sec: doping critical point}
\begin{figure}[t]
    \centering
    \includegraphics[width=1\linewidth]{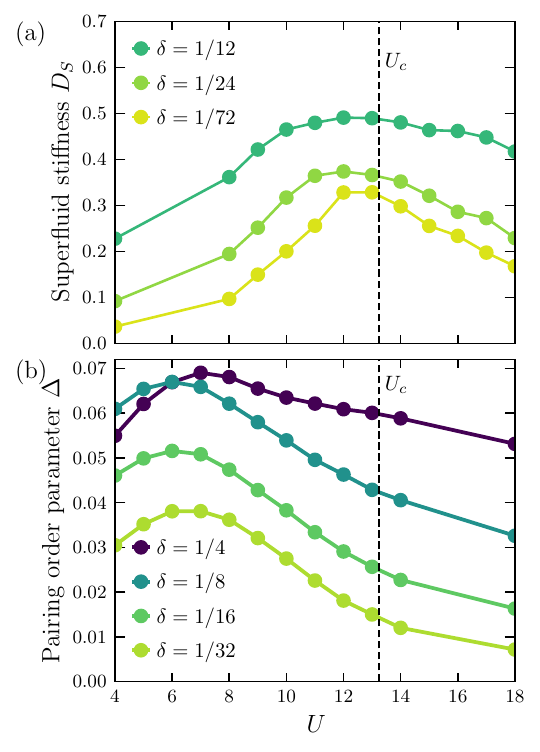}
    \caption{Doping the Hofstadter-Hubbard model around the critical point. (a) Superfluid stiffness $D_s$ as a function of $U$ for several different doping levels at $N=432$, computed from $d^2E/d\phi_y^2$. At low doping, the stiffness peaks around the critical point. (b) Pairing order parameter extracted from the square root of the average pair correlations beyond distance $4$ on the $192$ site cluster as a function of $U$ for several different doping levels. It peaks at small $U$ and is featureless around the critical point.}  
    \label{fig: main sc}
\end{figure}
We further characterize the pairing order parameter and superconducting stiffness. 
We compute the stiffness by threading flux $\phi$ through the torus, and computing $\frac{d^2E}{d\phi^2}$, as we detail in App.~\ref{sec: appendix stiffness}. The pairing order parameter is computed as $\sqrt{C_p(\v{r})}$ for $|\v{r}| > 4$.

We find that the stiffness is enhanced around the quantum critical point.
In Fig.~\ref{fig: main sc}(a), we show the superfluid stiffness for various $U$ and doping levels. From the Nelson-Kosterlitz criterion~\cite{nelson1977universal}, we estimate the transition temperature at $U=13, \delta=1/72$ to be $0.12t$, indicating a high transition temperature. At small doping, the stiffness is enhanced near the critical point of $U\sim 13$. The relative height of the peak diminishes as we increase doping.

A simple dimensional analysis can shed light on the peak of the stiffness~\cite{Divic2025PNAS}. The stiffness $D_s$ has units of energy$\sim (\text{length})^{-1}$. The doped problem has two length scales: $\delta^{-\frac{1}{2}}$ and the correlation length of the parent state $\xi$. Given $\xi \to \infty$ at the critical point, we can fix the scaling form of the stiffness to be $D_s = \sqrt{\delta} F\left(\frac{\sqrt{\delta}}{\xi}\right)$, where $F(0) = 1$, consistent with the scaling of the degeneracy temperature of free massless 2D boson gas. The functional form away from the critical point can be inferred from the degeneracy temperature of a massive 2D free boson gas, which goes as $\frac{\delta}{m}$. 
At small enough $\delta$, therefore, the stiffness peaks around the critical point.

In contrast, the pairing order parameter shows no feature at the quantum critical point. The order parameter peaks around $U=6-7$, away from the critical point. This is consistent with pre-formed pairs, arising either from binding of electrons in the IQH phase~\cite{Divic2025PNAS, Chen2026PRL} or from anyon superconductivity from semions~\cite{laughlin_relationship_1988,laughlin_superconducting_1988,fetter_random_phase_1989,ChenWilczekWittenHalperin1989,WenWilczekZee1989,WenZee1989,WenZee1990,lee_anyon_1989,HosotaniChakravarty}, whose mobility peaks at the critical point.

\PRLsec{Discussion}
Our results demonstrate the power of neural quantum states to understand the physics of topological quantum phase transitions. By accessing large system sizes with an expressive variational ansatz we uncovered properties of critical phases at scales beyond what is achievable with other numerical methods.

The systematic decrease of the $2e$ charge gap with system size and the power-law form of the charge structure factor are consistent with a conformally invariant quantum critical point. Our diagnostics, however, do not determine critical exponents, verify $z=1$ scaling, or match the operator content against candidate field theories~\cite{BarkeshliMcGreevy2014,Lee2018, zhou2025chern}. Developing techniques that extract such universal data from critical variational wavefunctions is an important direction for future work.

Our observation of an enhanced superfluid stiffness around the quantum critical point suggests critical charge fluctuations of the parent state stabilize the superconductor. Considering that the pairing order parameter does not show a similar enhancement, we posit that the critical fluctuations enhance superconductivity by promoting pair mobility. 
Surprisingly, despite the differences in how the superconductor forms upon doping the IQH and the CSL, the phenomenological behavior is remarkably similar on either side of the critical point. This suggests that doping near a quantum phase transition~\cite{Divic2025PNAS, pichler2026microscopic} may provide a more favorable route to strong superconductivity than doping a parent insulating phase.   

An experimental realization of this model may be achieved either in a cold-atom system~\cite{aidelsburger2013realization}, or in moir\'e systems~\cite{wu2018hubbard, kuhlenkamp2024chiral}. The high superconducting transition temperatures relative to the bare kinetic energy scale make this an ideal system for exploring the physics of strongly correlated superconductors.

\PRLsec{Data Availability}
Data used in the figures, as well as the variational energies of the underlying wavefunction, are available \href{https://github.com/chrisrothUT/HofstadterVariationalEnergies}{here}.

\PRLsec{Note added} Ref.~\cite{gu2026spincharge}, which appeared on the same day, also studies the triangular-lattice Hofstadter-Hubbard model using neural quantum states; the results are mutually consistent when they overlap.

\PRLsec{Acknowledgements}
We thank the authors of Ref.~\cite{gu2026spincharge} for coordinating arXiv submission.
We acknowledge discussions with Ashvin Vishwanath and T. Senthil. C.R. also acknowledges discussions with Ao Chen, Hannah Lange, and Antoine Georges. T.S. additionally thanks Stefan Divic, Valentin Cr\'epel, Clemens Kuhlenkamp, Xueyang Song, and Mike Zaletel for previous collaborations on related topics.
The work of A.J.M. was supported in part by the National Science Foundation (NSF) MRSEC program through the Center for Precision-Assembled Quantum Materials (PAQM) under Grant Number DMR-2011738.
This work was partly supported by JST PRESTO, Japan, Grant Number JPMJPR2455 to T.S.
Figure 1 was created by Lucy Reading-Ikkanda. 
Simulations were performed using the Scientific Computing Core at the Flatiron Institute. 
The Flatiron Institute is a division of the Simons Foundation.

\bibliographystyle{unsrt}
\bibliography{references}

\onecolumngrid
\appendix
\newpage

\section{Geometry and gauge choice} \label{sec: appendix geometry+gauge}

\begin{figure*}[t]
    \centering
    \includegraphics[width=0.3\linewidth]{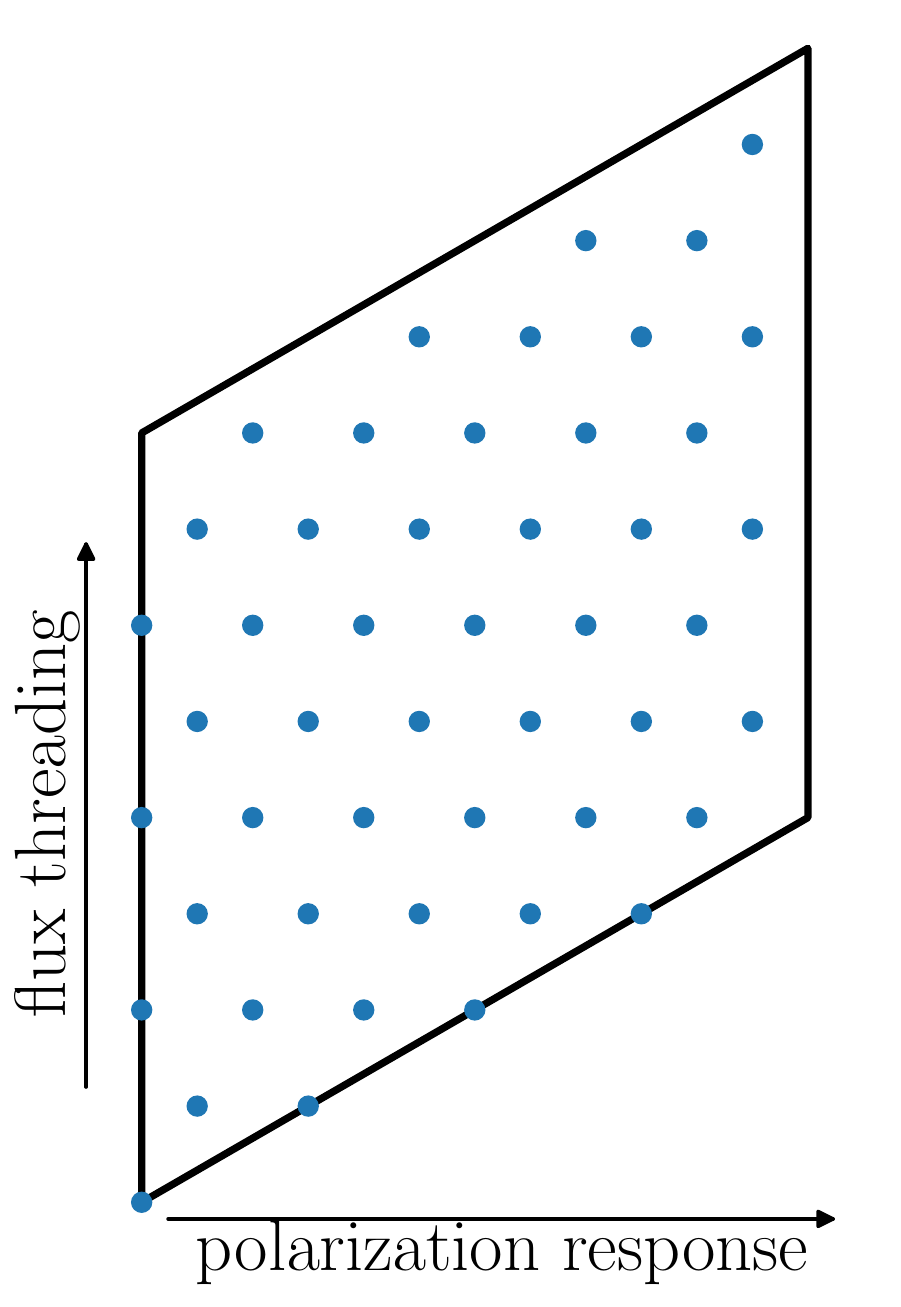}
    \caption{Diagram of the tori simulated in this work. Here we show a $\sqrt{3}L \times \sqrt{3}L$ cluster with $L=4$ ($N=48$ sites) as an example. In the main text we discuss $L=8,10,12$ with $N=192,300,432$. In the flux threading experiments, flux is always applied along the $y$-axis, which is the shortest loop around the torus. For Hall-response calculations, we measure the polarization response in the orthogonal $x$-direction.} 
    \label{fig: flux threading + torus}
\end{figure*}

Our simulations are performed on $\sqrt{3}L \times \sqrt{3}L$ lattices on a torus geometry shown in Fig.~\ref{fig: flux threading + torus} as done in Refs.~\cite{wietek2024quantum, roth2023high}, for $L=8, 10, 12$. This can be constructed starting from an $L \times L$ triangular lattice on a torus with lattice vectors $\{3/2, \sqrt{3}/2\}$ and $\{0, \sqrt{3}\}$. From this, the $\sqrt{3}L \times \sqrt{3}L$ lattice is formed by placing three sites in each unit cell, displaced from the unit-cell origin by $\{0,0\}$, $\{1/2, \sqrt{3}/2 \}$, and $\{1, \sqrt{3} \}$. This lattice maintains $D_6$ symmetry, and the shortest loops around the torus now traverse the next-nearest neighbor direction.

\begin{figure}
    \centering
    \includegraphics[width=0.6\linewidth]{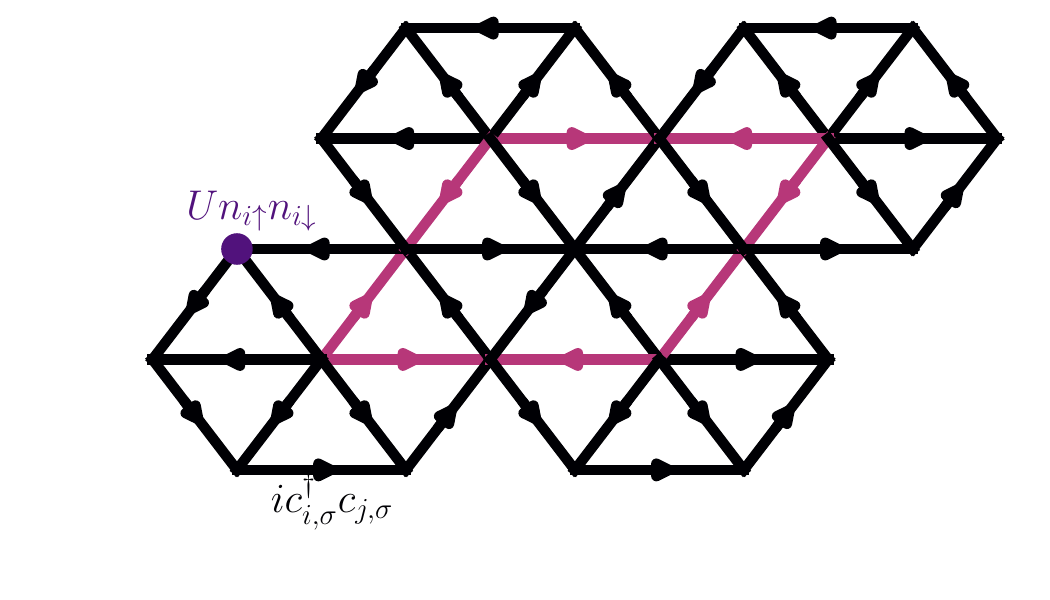}
    \caption{Diagram of the Hofstadter-Hubbard Hamiltonian at $\pi/2$ flux. The signs of the imaginary hoppings are indicated by the direction of the bonds. Each unit cell contains twelve bonds, which can be drawn as a hexagon with a $C_6$ symmetric point at the center. Here we show a $2 \times 2$ tiling of these bond hexagons, where the magnetic unit cell is drawn in magenta. The order parameter is defined on this same hexagon, as shown in the inset to Fig.~\ref{fig: topoSC}(a).}
    \label{fig:ham hoppings}
\end{figure}

We choose a $C_6$ symmetric gauge as done in \cite{Divic2025PNAS,kuhlenkamp2024chiral}, which has a $2 \times 2$ unit cell with 12 bonds. A diagram of the Hamiltonian is shown in Fig.~\ref{fig:ham hoppings}. The arrows represent kinetic hopping of $it$. Due to hermiticity the hopping sign is $-it$ in the opposite direction of the arrows. This unit cell has 4 sites and 12 bonds, and is therefore tiled $3L^2/4$ times in order to form the $\sqrt{3}L \times \sqrt{3}L$ lattice. 

\section{Identifying the IQH and CSL phases from flux threading} \label{sec: appendix IQH-CSL}

The IQH and CSL can be distinguished by their response to flux threading, which measures the Hall response to an applied field.

\begin{figure*}[t]
    \centering
    \includegraphics[width=0.9\linewidth]{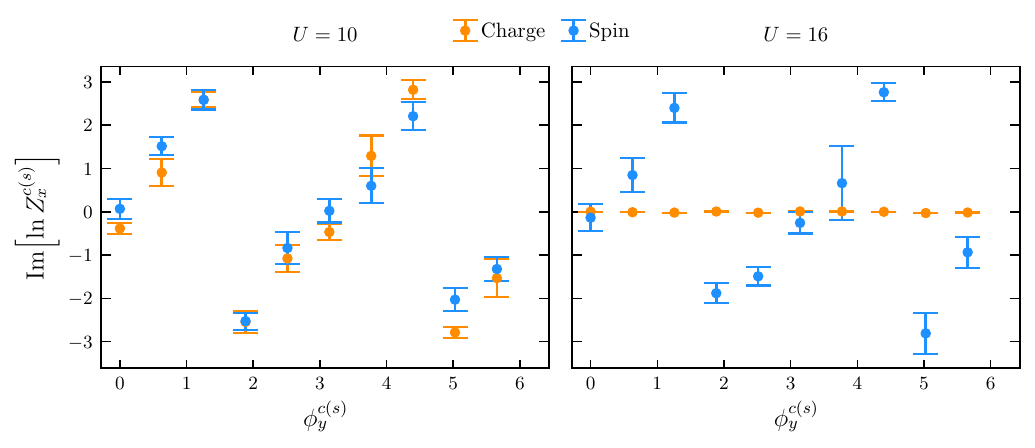}
        \caption{Identifying phases of the $\pi/2$ Hofstadter-Hubbard model at half filling. Charge and spin quantum Hall response at $U=10$ (left) and $U=16$ (right) on the 192 site cluster. The flux threading and measurement directions are shown in Fig.~\ref{fig: flux threading + torus}. As expected, there is two-fold winding of both the charge and spin polarization in the IQH phase, while only the spin polarization winds in the CSL phase. 
        }
    \label{fig: IQH-CSL}
\end{figure*}

Our real space torus is spanned by two basis vectors
\begin{equation}
    \v{L}_1 = L(0, \sqrt{3}), \quad \v{L}_2 = L\left(\frac{3}{2}, \frac{\sqrt{3}}{2}\right).
\end{equation}

We thread the flux in the $y$ direction:

\begin{equation}
    e\v{A} = \left(0, \Phi_y \frac{1}{\sqrt{3}L}\right) \implies \v{L}_1 \cdot e\v{A} = \Phi_y, \quad \v{L}_2 \cdot e\v{A} = \frac{\Phi_y}{2}.
\end{equation}

Many-body polarization is measured using the phase of the operator $\exp[i\v{G} \cdot \hat{\v{R}}]$, where $\v{G}$ is a reciprocal lattice vector of the torus and $\hat{\v{R}}$ is the position operator. For our torus, the reciprocal lattice is spanned by the basis vectors
\begin{equation}
    \v{G}_1 = \frac{4\pi}{3\sqrt{3}L}\left(\frac{\sqrt{3}}{2}, -\frac{3}{2}\right), \v{G}_2 = \frac{4\pi}{3 L}(1, 0).
\end{equation}
We measure the many-body polarization using $\v{G}_2$:
\begin{equation}
    P_x = \frac{3L}{4\pi} \Im \log \braket{\exp\left[\frac{4\pi i}{3L} \hat{X}\right]} \equiv \frac{3L}{4\pi} \Im \log \braket{Z_x},
\end{equation}
where $\hat{X} = \sum_{\v{r}} r_x \hat{n}_{\v{r}}$.

Let us now compute the Hall conductance. We first compute the current density from the time derivative of the polarization:

\begin{equation}
    j_x = \frac{1}{\mathrm{Area}} \frac{dP_x}{dt} = \frac{2}{3\sqrt{3}L^2}\frac{3L}{4\pi}\frac{d(\Im \log \braket{\exp\left[\frac{4\pi i}{3L} \hat{X}\right]})}{d\Phi_y} \frac{d\Phi_y}{dt} = \frac{1}{\sqrt{3}L}W \times \frac{d\Phi_y}{dt}
\end{equation}
where $W = (P_x(\Phi_y = 2\pi) - P_x(\Phi_y=0))/2\pi$ is the winding number of the threading process. On the other hand, the electric field is given by $\v{E} = \frac{d\v{A}}{dt} = \frac{1}{\sqrt{3}L} \frac{d\Phi_y}{dt}$. Combining these, we get
\begin{equation}
    j_x = WE \implies \sigma = \frac{e^2}{h} W.
\end{equation}

We can similarly measure response to a spin flux by writing

\begin{equation}
    e\v{A}_s^\sigma = \sigma\left(0, \Phi_y \frac{1}{\sqrt{3}L}\right), \quad P_x^\sigma = \frac{3L}{4\pi} \Im \log \braket{\exp\left[\frac{4\pi i}{3L}(\hat{X}_\uparrow - \hat{X}_\downarrow)\right]} 
\equiv
\frac{3L}{4\pi} \Im \log \braket{Z^{\rm spin}_x}.
\end{equation}

The same algebra as for the charge Hall conductivity gives us
\begin{equation}
    \sigma_{xy}^s = \frac{\hbar}{8\pi} W^s.
\end{equation}

In Fig.~\ref{fig: IQH-CSL} we show the Hall response to charge and spin flux for $U=10$ and $U=16$. For $U < U_c$ we have two copies of an IQH formed by the up and down spins, so under flux threading the up and down spins are moved independently. The charge and spin quantum Hall responses should then show winding numbers $W = W^s = 2$, as is clearly observed at $U = 10$. 

In contrast, for $U > U_c$ we expect that the charge is localized, and we should no longer see a winding of the charge many-body polarization under flux threading. As expected for the CSL phase, we only see a spin response at $U=16$. 

\section{Detecting quantum criticality through structure factors} \label{sec: structure factors}

\subsection{Structure factor and quantum criticality}
For an operator $\hat{O}$, we define the connected spatial correlation function as follows: 

\begin{equation}
C_{\hat O}({\bf r}) = \frac{1}{N} \sum_{\bf x} \Big[ \langle {\hat O}^\dagger_{\bf x + \bf r} {\hat O_{\bf x}} \rangle- \langle {\hat O}_{\bf x + \bf r} \rangle  \langle {\hat O}_{\bf x} \rangle  \Big].
\end{equation}

For phases with long range order, $C_{\hat O}({\bf r})$ should be non-zero as $|{\bf r}| \rightarrow \infty$, while in most other cases it falls off exponentially. The exception is quantum critical phases where $C_{\hat O}({\bf r}) \sim 1/|{\bf r}|^\eta$ at large $|{\bf r}|$. In order to distinguish a power law from an exponential, one must determine whether $\log(C_{\hat O}({\bf r}))$ decays linearly or super-linearly with $\log(|{\bf r}|)$. However, in practice it is hard to fit a long-range correlation tail from Monte Carlo data, as the statistical noise dominates at large $|{\bf r}|$ and causes the sign of the correlation function to oscillate.

To ameliorate this problem, we consider correlation functions in momentum space, often called structure factors,     
\begin{equation}
C_{\hat O}({\bf k}) =  \sum_{\bf r} e^{i {\bf k} \cdot {\bf r}} C_{\hat O}({\bf r}).
\end{equation}

For each value of $|{\bf k}|$, $C_{\hat O}({\bf k})$ contains a sum over $N$ terms which improves the statistical signal by $\sqrt{N}$---this improves the resolution by a factor of 10--20 on the system sizes we study. Furthermore, $C_{\hat O}({\bf k})$ is positive semidefinite, and thus taking the logarithm is unproblematic. 

For a phase with long range order, $C_{\hat O}({\bf k})$ should scale extensively with system size. On the other hand, for exponentially decaying correlations, $C_{\hat O}({\bf k})$ should be analytic at $|{\bf k}| = 0$ and therefore must scale quadratically at small $|{\bf k}|$: $C_{\hat O}({\bf k}) - C_{\hat O}({\bf k} = 0) \sim |{\bf k}|^2$.

However, if $C_{\hat O}({\bf r})$ follows a power law with $\eta > 2$, the structure factor can be non-analytic at ${\bf k}=0$, where the small-$|{\bf k}|$ scaling goes as $C_{\hat O}({\bf k}) \sim \alpha |{\bf k}|^2 + \beta|{\bf k}|^{\eta-2}$. Therefore, by examining the small-$|{\bf k}|$ behavior of the structure factor we can identify critical fluctuations.

\subsection{Kinetic energy fluctuations}

\begin{figure*}[t]
    \centering
    \includegraphics[width=0.9\linewidth]{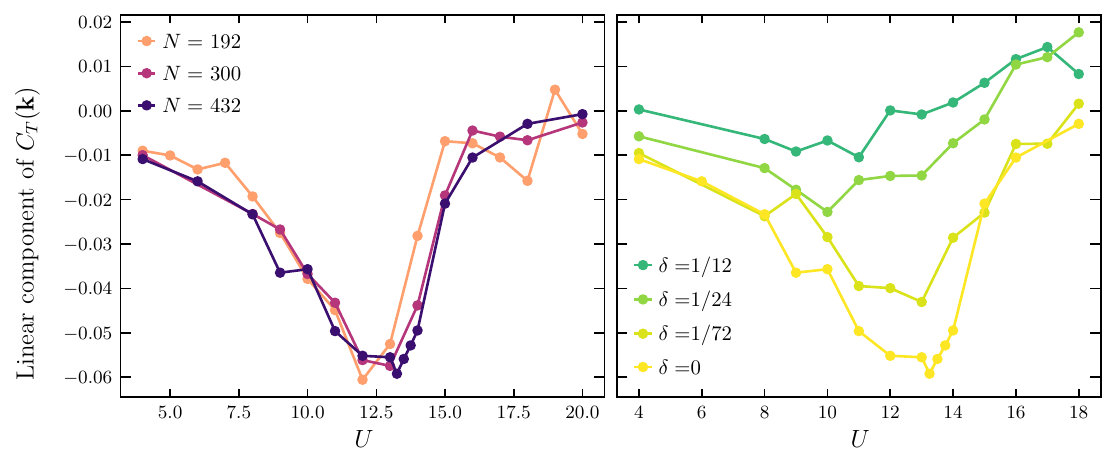}
    \caption{Behavior of the kinetic energy structure factors at small momenta. The linear component of the kinetic energy structure factor, $c_1$, is extracted from a quadratic fit of $C_{\hat T}({\bf k})$ for $|{\bf k}| < 2\pi/3$. The system size and $U$ dependence of $c_1$ is plotted on the left, while the doping dependence (for $N=432$) is plotted on the right. There are anomalous kinetic energy fluctuations around the critical point, which gradually dissipate with doping.}
    \label{fig: energy structure}
\end{figure*}

In the main text Fig.~\ref{fig: main QCP} we present evidence that there are critical charge fluctuations in the vicinity of the QCP. Here we show that there are critical kinetic energy fluctuations as well, which persist to finite doping. First, we define the local kinetic energy operator,
\begin{equation}
    {\hat T}_{\bf x} = i \ \textrm{sign}(\boldsymbol{\delta}) \sum_{\sigma = \uparrow, \downarrow} ({\hat c^\dagger}_{\bf x + \boldsymbol{\delta}/2,\sigma} {\hat c}_{\bf x - \boldsymbol{\delta}/2,\sigma} - {\hat c^\dagger}_{\bf x - \boldsymbol{\delta}/2,\sigma} {\hat c}_{\bf x + \boldsymbol{\delta}/2,\sigma}),
\end{equation}
where $\boldsymbol{\delta}$ is a nearest-neighbor lattice displacement and $\textrm{sign}(\boldsymbol{\delta})$ are the signs shown in Fig.~\ref{fig:ham hoppings}. Again we look at the small-$|{\bf k}|$ behavior of the structure factor $C_{\hat T}({\bf k})$. Unlike the charge structure factor, here $C_{\hat T}({\bf k} = {\bf 0}) \ne 0$. Therefore we find it is best to fit the small-$|{\bf k}|$ behavior with a quadratic function, $C_{\hat T}({\bf k}) \sim c_0 + c_1 |{\bf k}| + c_2 |{\bf k}|^2$. Any linear component indicates that $C_{\hat T}({\bf k})$ is non-analytic at ${\bf k} = 0$. In the left panel of Fig.~\ref{fig: energy structure}, we plot the linear coefficient of the fit, $c_1$, as a function of $U$ at half filling. This component is zero deep in the insulator phase but becomes strongly negative around the critical point, signaling critical fluctuations. In the right panel of  Fig.~\ref{fig: energy structure}, the behavior of the energy structure-factor fit at finite doping is shown. The peak in the $|{\bf k}|$-linear component decreases as a function of doping until it is undetectable at $\delta \sim 0.1$. This, in conjunction with the superfluid stiffness in main text Fig.~\ref{fig: main sc}, gives evidence that the critical point remains influential until about $10 \%$ doping. 

\subsection{Fits to structure factors}

\begin{figure*}
    \centering
    \includegraphics[width=0.9\linewidth]{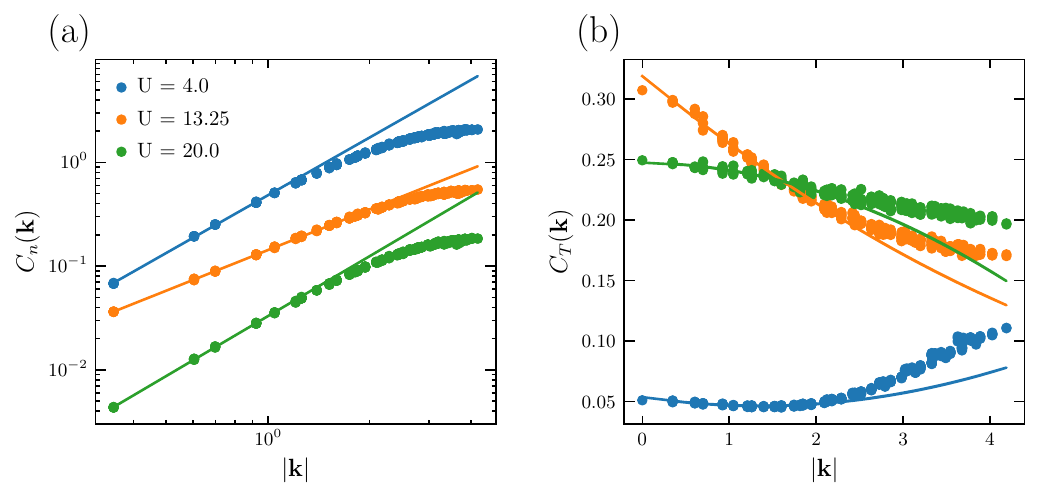}
    \caption{Fitting the small-$|{\bf k}|$ behavior of the structure factor for $432$ sites at half filling. Here we choose representative values of $U=4,13.25,20$ to show the fit for the IQH, QCP, and CSL. (a) Power law fit of the charge structure factor on a log-log plot. (b) Quadratic fit to the energy structure factor.}
    \label{fig:struct fits}
\end{figure*}

In this section we show details of the fits to the small $|{\bf k}|$ behavior of the charge and energy structure factors. The charge structure factor is $0$ at ${\bf k} = {\bf 0}$ and positive definite, therefore we can fit the low-$|{\bf k}|$ behavior to a power law. The results for $U=4$, which is deep in the IQH phase, $U=13.25$, which is near the critical point, and $U=20$, which is deep in the CSL phase, are shown in Fig.~\ref{fig:struct fits}(a). For these calculations we fit $|{\bf k}| < \pi/3$, as beyond these values the higher-order contributions to the structure factors of the insulators become relevant. However, as seen by the orange curve, the fit around the critical point persists to much larger $|{\bf k}|$.

Fitting the kinetic energy fluctuations is trickier as $C_{\hat T}({\bf k} = 0)$ is generically non-zero as it represents the variance of the kinetic energy \footnote{To avoid confusion, we note that the variance of the {\it total} energy is zero for an eigenstate but the variance of the {\it kinetic} energy can still be finite.}. Therefore, we would like to fit the data to the functional form $C_{\hat T}({\bf k}) = \alpha |{\bf k}|^2 + \beta |{\bf k}|^{\eta - 2} + \gamma$. In practice we find that fixing $\eta = 3$ works quite well. The fits to the energy structure factor for the IQH, QCP, and CSL are shown in Fig.~\ref{fig:struct fits}(b). As seen by the orange curve, there is sub-quadratic behavior at $U=13.25$ although there is a slight deviation from linear behavior at small-$|{\bf k}|$ points.

\section{Superfluid stiffness} \label{sec: appendix stiffness}

\subsection{Extraction of superfluid stiffness and BKT transition temperature}
We compute the superconducting stiffness from flux response. As defined in App.~\ref{sec: appendix IQH-CSL}, the
flux $\phi_y$ enters as a uniform vector potential $e\v{A} = \left(0, \frac{\phi_y}{\sqrt{3}L}\right)$. The superfluid weight in the $y$ direction is
the curvature of the ground-state energy per unit area with respect to
$A_y$~\cite{scalapino1993insulator},
\begin{equation}
    D_s = \frac{1}{\mathcal{A}} \frac{\partial^2 E}{\partial A_y^2}
    \bigg|_{A_y = 0},
\end{equation}
where $\mathcal{A} = |\v{L}_1 \times \v{L}_2| = \frac{3\sqrt{3}}{2}L^2$ is the
area of the torus. Using $\frac{\partial^2 E}{\partial \phi_y^2} =
\frac{1}{3L^2}\frac{\partial^2 E}{\partial A_y^2}$, we obtain
\begin{equation} \label{eqn: stiffness conversion}
    D_s = \frac{2}{\sqrt{3}} \frac{d^2 E}{d\phi_y^2}.
\end{equation}

Since the condensate carries charge $2e$, the phase stiffness of the pairing
order parameter is $\rho_s^{\rm pair} = D_s/4$, and the Nelson-Kosterlitz universal jump
condition~\cite{nelson1977universal} $\rho_s^{\rm pair}(T_{\rm BKT}) = \frac{2}{\pi} k_B T_{\rm BKT}$
yields an upper bound on the transition temperature,
\begin{equation}
    k_B T_{\rm BKT} \lesssim \frac{\pi}{2}\rho_s^{\rm pair}
    = \frac{\pi}{8} D_s.
\end{equation}

\subsection{Extracting stiffness from finite difference}

Assuming that the ground-state energy is quadratic in flux near $\phi_y=0$, Eq.~\eqref{eqn: stiffness conversion} may be approximated using a finite difference method on the larger systems,
\begin{equation} \label{eqn: stiffness finite difference}
D_s = \frac{4}{\sqrt{3}} \frac{E(\phi_y = \phi_s) - E(\phi_y = 0)}{\phi_s^2}, 
\end{equation}
where we set $\phi_s = 0.4 \times 2\pi $. This greatly reduces computational time as it only involves training the model at two flux values.

\begin{figure*}
    \centering
    \includegraphics[width=0.9\linewidth]{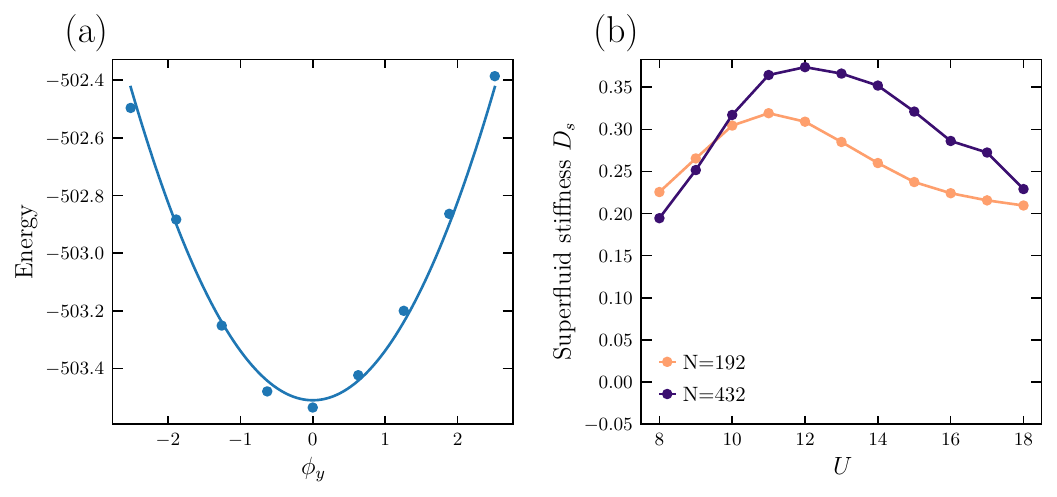}
    \caption{(a) Energy as a function of flux threaded on the 192 site cluster at $U=8$ and $\delta = 1/8$. A quadratic fit is shown, indicating that $E(\phi_y) \sim \phi_y^2$ to good approximation. (b) Superfluid stiffness as a function of $U$ on the $192$ and $432$ site clusters at $\delta = 1/24$. As the system size is increased, the stiffness is slightly enhanced in the vicinity of the critical point.}
    \label{fig:stiffness fit}
\end{figure*}

To verify the accuracy of this method, we confirmed that $E(\phi_y)$ is a quadratic function, and that Eq.~\eqref{eqn: stiffness finite difference} reproduces a quadratic fit to good accuracy. In Fig.~\ref{fig:stiffness fit} we show a fit to $E(\phi_y)$ as a function of $\phi_y$ at $U=8, \delta = 1/8$, showing a clear quadratic behavior. In Table~\ref{tab: stiffness comparison}, we compare the stiffness extracted via the quadratic fit on $\phi_y = 2\pi \times \{0,0.1,0.2,0.3,0.4\}$ versus Eq.~\eqref{eqn: stiffness finite difference} at representative points corresponding to doping the IQH, QCP, and CSL. The two calculations show good agreement.

\begin{table}[]
    \centering
    \begin{tabular}{c|c|c|c}
        & $U=4$ & $U=12$ & $U=20$ \\ \hline
       Quadratic fit  & 0.0815 & 0.313 & 0.149  \\ \hline
       Finite difference  & 0.0813 & 0.321 & 0.148 
    \end{tabular}
    \caption{Extraction of superfluid stiffness using two different methods, a quadratic fit as a function of $\phi_y$ and the finite difference method in Eq.~\eqref{eqn: stiffness finite difference} for $\delta = 1/32$ on the 192 site cluster.} 
    \label{tab: stiffness comparison}
\end{table}
    
\subsection{Finite size effects of superfluid stiffness}

In Fig.~\ref{fig:stiffness fit}(b) the superfluid stiffness is shown on the $192$ and $432$ site clusters at $\delta = 1/24$ as a function of $U$. Away from $U_c$ the superfluid stiffness is approximately size invariant, but around the critical point the stiffness has a larger peak for the $N=432$ site system. This is consistent with the energy gap, which decreases and moves to larger $U$ as system size is increased.  

\section{Details of variational optimization}

\subsection{Hyperparameters and training procedure} \label{sec: appendix hyperparameters}

For all of the calculations a hidden fermion Pfaffian state wavefunction \cite{chen2025neural} was used in conjunction with a residual convolutional neural network (R-CNN) with $36$ features, $8$ layers and $8$ hidden fermions. The visible-visible couplings have a $2 \times 2$ sublattice structure such that they have the same unit cell as the Hamiltonian couplings. 

A separate calculation was done for each doping level, system size, and flux threading, while we used transfer learning to compute the ground state as $U$ is varied. We first trained at $U=4$ before adiabatically increasing $U$ and making sure the calculation was converged at each step. 

First, we optimized the mean field on the Hubbard model at $U=3$ with a pinning field consistent with the order parameter shown in the inset of Fig.~\ref{fig: topoSC}(a).

\begin{equation}
    H_{\rm mf}  = it\sum_{i,j} \tau_{i, j} c^\dagger_i c_j + \sum_i U_i n_{i\uparrow} n_{i\downarrow} + \mu \sum_{i,\sigma} n_{i, \sigma} + \Delta \sum_{\langle \alpha,\beta \rangle} \textrm{phase}(\alpha,\beta) (c_{\alpha,\uparrow}  c_{\beta,\downarrow} - c_{\alpha,\downarrow}  c_{\beta,\uparrow}) 
    \label{eq: mean field hamiltonian}
\end{equation}
as done in Ref.~\cite{roth2025superconductivity}. For all calculations we use $\Delta = 0.2$. We find that adding the pinning field to the mean field starting point substantially improves the convergence as well as the energy after a fixed number of iterations. We show the energy after $1000$ training iterations for two different doping levels at $\Delta = 0, 0.2$ in Table~\ref{tab: pinning field}.

\begin{table}[]
    \centering
    \begin{tabular}{c|c|c}
     &  $\Delta = 0$  & $\Delta = 0.2$ \\ \hline
     $\delta = 1/16$ & -530.26(2)  & -531.99(2) \\ \hline
     $\delta = 1/8$ & -501.91(2) & -502.63(2)
    \end{tabular}
    \caption{Energy with and without pinning field after $1000$ training iterations at $U=8$ on the 192 site cluster for two different doping levels.}
    \label{tab: pinning field}
\end{table}
We find that after $1000$ iterations the wavefunction without pinning has still not found the correct superconducting order, explaining the much higher energy. 

\subsection{Zero-variance extrapolation} \label{sec: appendix zero-variance}

Throughout this work we compute gaps by considering the difference in energy between two eigenstates. Furthermore, the superfluid stiffness is computed from the dependence of the ground-state energy on the threaded flux. Thus, the error in these results is set by the error scale of the variational energies. 

In order to estimate the error in the variational energy, we perform a zero-variance extrapolation \cite{fu2023variance, chen2024empowering}, using two states: the raw wavefunction computed using a CNN, and the wavefunction after $C_6$ + spin-inversion projection.
\begin{equation}
\langle {\bf n} |\psi_{\rm symm}\rangle = \sum_{g \in G} \chi_g \langle g  {\bf n} |\psi   \rangle   
\end{equation}
where $G$ is the group containing $C_6$ rotations as well as a spin inversion operation, and $g {\bf n}$ is the transformation of the Fock state ${\bf n}$ by $g$. Here we find the lowest possible $|\psi_{\rm symm} \rangle$ by trying all possible characters $\chi_g$. For the doping levels we consider, where the number of holes is always even, the eigenvalue under a $\pi/3$ rotation is $\exp \Big(i\frac{\pi}{3}\big(4 + \frac{N-N_e}{2}\big)\Big)$, while under spin parity it is $\exp \Big(i \pi\frac{N-N_e}{2} \Big)$. Once we have obtained $|\psi\rangle$ and $|\psi_{\rm symm}\rangle$ the extrapolated energy is computed as 
\begin{equation}
E_{\rm ext} = E_{\rm symm} + \frac{E_{\rm symm} - E}{\sigma^2_{\rm symm} - \sigma^2} \sigma^2_{\rm symm}.
\end{equation}

\begin{figure*}
    \centering
    \includegraphics[width=0.9\linewidth]{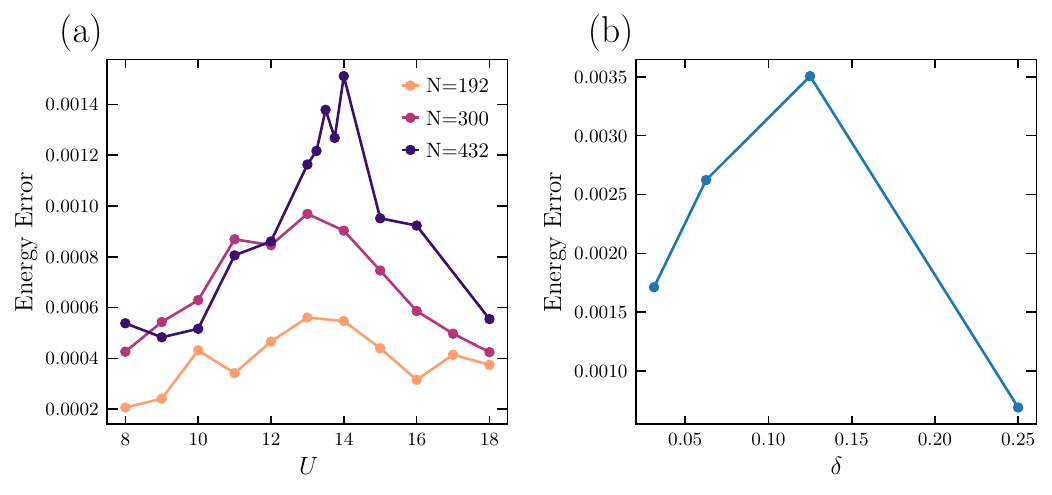}
    \caption{Estimates of variational accuracy. The error is estimated from the difference between the variational energy after $C_6$ and spin parity projection and the zero-variance extrapolation described in App.~\ref{sec: appendix zero-variance}. (a) Energy error per site as a function of $U$ at half filling for the three different system sizes. (b) Energy error per site as a function of doping at $U=8$ on the $192$ site cluster.}
    \label{fig:energy error half}
\end{figure*}

In Fig.~\ref{fig:energy error half} we show approximate energy errors using this extrapolation procedure. Here we see that the energy errors are fairly small both at half filling and with hole doping.    

\subsection{Numerical Resources}

The data shown in this paper required approximately $10^{5}$ hours on NVIDIA H200 GPUs. In Table~\ref{tab: GPU resources} the approximate time needed for $2000$ training iterations at half filling is shown as a function of system size.  

\begin{table}[]
    \centering
    \begin{tabular}{c|c|c|c}
     $N$  & 192  & 300 & 432 \\ \hline
     GPU Hours & 150 & 400 & 1000 \\ 
    \end{tabular}
    \caption{Approximate H200 GPU hours needed to train models with $N$ sites at half filling for $2000$ iterations.}
    \label{tab: GPU resources}
\end{table}

Upon hole doping, the computation becomes slightly cheaper due to the smaller Pfaffian in the output layer. 

\end{document}